\documentclass[aps,prl,preprintnumbers,twocolumn,amsmath,amssymb]{revtex4-1} 
\usepackage[utf8]{inputenc}
\usepackage[english]{babel}
\usepackage{graphicx}
\usepackage{subfiles}
\usepackage{blindtext}
\usepackage{amssymb}
\usepackage{amsmath}
\usepackage{graphicx}
\usepackage{float}
\usepackage[braket, qm]{qcircuit}
\usepackage{tikz}
\usepackage{float}
\usepackage{appendix}
\usepackage{wrapfig}
\usepackage{xargs}      
\usepackage{xcolor}
\usepackage[colorlinks]{hyperref}
\usepackage{cleveref}
\definecolor{winered}{rgb}{0.8,0,0}
\definecolor{darkb}{rgb}{0,0,0.8}
\hypersetup{
    colorlinks=true,
    citecolor=blue,
    linkcolor=winered,
    filecolor=red,      
    urlcolor=darkb,
}
\usepackage[margin=1in]{geometry}
\newcommand{\dket}[1]{|#1 \rangle}
\newcommand{\dbra}[1]{\langle #1 |}

\begin{document}
\bibliographystyle{apsrev}

\title{Real-time quantum calculations of phase shifts using wave packet time delays 
}
\author{Erik Gustafson$^1$}
\author{Yingyue Zhu$^2$}
\author{Patrick Dreher$^3$ $^4$}
\author{Norbert M. Linke$^2$}
\author{Y. Meurice$^1$}
\affiliation{$^1$ Department of Physics and Astronomy, The University of Iowa, Iowa City, IA 52242, USA }
\affiliation{$^2$ Joint Quantum Institute and Department of Physics, University of Maryland, College Park 20740, USA }
\affiliation{$^3$ Department of Computer Science, North Carolina State University, Raleigh, NC  27695, USA}
\affiliation{$^4$ Department of Physics, North Carolina State University, Raleigh, NC 27695, USA}
\date{\today}
\begin{abstract}
We present a method to extract the phase shift of a scattering process using the real-time evolution
in the early and intermediate stages of the collision in order to estimate the time delay of a wave packet. This procedure is convenient when using noisy quantum computers for which the asymptotic out-state behavior is unreachable. 
We demonstrate that the challenging Fourier transforms involved in the state preparation and measurements can be implemented in $1+1$ dimensions with current trapped ion devices and IBM quantum computers. We compare quantum computation of the time delays obtained in the one-particle quantum mechanics limit and the 
scalable quantum field theory formulation with accurate numerical results. We discuss the finite volume effects in the Wigner formula connecting time delays to phase shifts. The results reported involve two- and four-qubit calculations, and we discuss the possibility of larger scale computations in the near future. 
\end{abstract}
\maketitle

{\it 1. Introduction.} The idea of simulating quantum field theory with quantum computers has gained considerable interest recently \cite{PreskillPOS}. In the context of high-energy and nuclear physics, a long term motivation is to develop quantum computing methods that perform real-time evolution for  lattice quantum chromodynamics (QCD).  This is an ab-initio, ultraviolet complete, theory of strong interactions which has been very successful in describing the static properties of hadrons and nuclei \footnote{See review 17 of the PDG \cite{pdg}.}. Importance sampling methods, used successfully today for lattice QCD at Euclidean time, are not effective for dealing with the rapid oscillations of real-time unitary operators acting on large Hilbert spaces.  Currently, physicists resort to semi-empirical algorithms such as Pythia and Herwig \cite{pythiamanual,HerwigManual}) to interpret hadronic collider data. Doing ab-initio calculations for jet physics would represent a major accomplishment and remains a long-term goal for the high-energy community and strategies to deal with parton distributions are outlined in Refs. \cite{PhysRevD.102.016007, Lamm_2020}. Related methods for out-of-equilibrium processes in many-electron systems and information paradoxes in quantum gravity \cite{PreskillMicrosoft} would also have a large potential impact. 

Quantum computers offer an alternative solution to the sign problem plaguing current efforts. A sensible approach would be to follow the sequence of models that has been successful for the development of lattice QCD at Euclidean time on classical computers \cite{RevModPhys.51.659,RevModPhys.55.775}. The first step in this sequence is to study and understand the properties and behavior of the quantum Ising model (QIM) on today's Noisy Intermediate-Scale Quantum (NISQ) hardware.  

Real time evolution involving a limited number of sites for the QIM has already been attempted using a few qubits on gate based quantum computers \cite{CerveraLierta2018exactisingmodel,Lamm:2018siq,GustafsonIsing,gustafson2019real, Kim_2020,yeteraydeniz2021scattering,vovrosh2020confinement,Kandala:2017aa,Kandala_2019,Salath__2015,Labuhn_2016,2017Natur.551..601Z,PhysRevE.58.5355,2017Natur.551..579B}, as well as developments in progress for more complicated models \cite{Alexandru_2019,Zohar_2013, Martinez_2016, Klco:2018kyo,brower2020lattice}. Today researchers must grapple with discrepancies between the measured qubit-state occupations on quantum platforms and the exact evolution after only a few Trotter steps.  Currently, error-mitigation methods such as zero-point extrapolation \cite{Richardson,Klco:2018kyo} are necessary to improve the computational results. It has been shown \cite{gustafson2019real} that by modelling four qubits on an IBM Q quantum computing hardware platform these mitigation methods provide a reasonable extrapolation for times of the order of the approximate periodicity of the problems considered. It is possible to use Trotter steps that are more than 20 times larger than rigorous bounds in $(\delta t)^2$ or $(\delta t)^3$ would naively suggest, without encountering major discretization errors \cite{gustafson2019real,ybook}. This allows us to reach larger time scales for studying scattering processes.

\begin{figure}[h]
\centering
\includegraphics[width=0.47\textwidth]{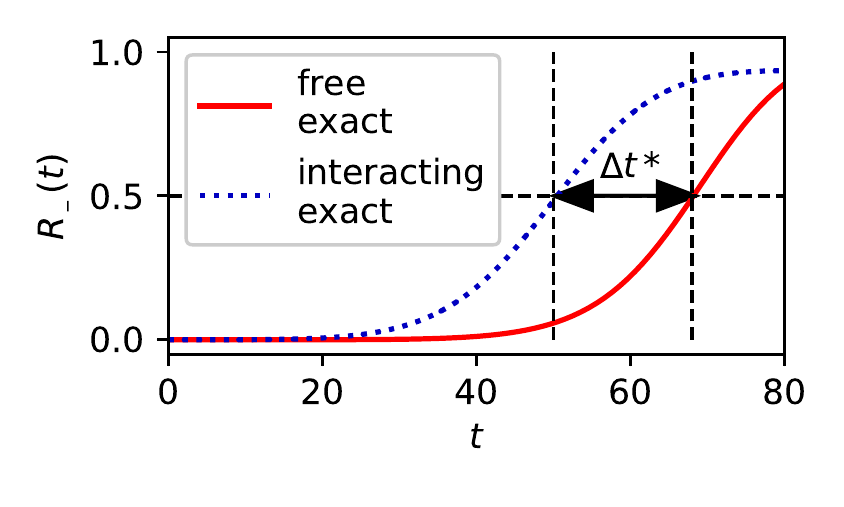}
\caption{Illustration of the measurement of the time delay between the free and interacting wave packets. The renormalized reflection probability $R_-(t)$ is defined in Eq. (\ref{eq:rminus})}
\label{fig:diagram}
\end{figure}

In this Letter, we demonstrate that it is possible to use state-of-the-art NISQ devices to prepare and evolve suitable wave packets for the QIM. We show that it is possible to project the wave-function in the early stages of a collision process onto momentum states and to pinpoint a time that corresponds to the middle of the collision. By introducing an extra interaction, we obtain a time delay $\Delta t^\star$ illustrated in Fig. \ref{fig:diagram} that is half of the time delay $\Delta t_W$ invoked in Wigner formula \cite{PhysRev.98.145} given in Eq. (\ref{eq:we}) to estimate the derivative of the phase shift with respect to the momentum. 

Phase shifts are a key measurement in the scattering process and represent the total change of phase due to interactions. Significant progress has been made in calculating them from lattice QCD in Euclidean time \cite{Luscher1986,Briceno:2013bda, Briceno:2016mjc,Briceno:2017max, SavageLAT2016,Davoudi:2018wgb}.  In standard textbooks, phase shifts and scattering amplitudes are estimated from asymptotic data long after the collision processes have occurred. However, for NISQ devices with limited coherence time or gate-depth, using the information from the early stages of the collision is advantageous.  We show that this idea can be implemented on both a quantum computer using superconducting transmon qubits and a trapped ion system operating at the University of Maryland \cite{Debnath2016}.

The article is organized as follows. We introduce the quantum Ising model with an extra interaction and its Hilbert space.    We show that it is possible to implement the three steps of the calculation of the S-matrix elements: 1) preparation of the initial state, 2) real-time evolution, and 3) measurement of the probability for a particular final state. We then extract the phase shift by comparing the cases with and without an external potential.  Steps 1) and 3) involve Fourier transforms and are very challenging with NISQ devices. This is why we first restrict ourselves to the quantum mechanics limit where one-particle states interact with an external potential localized at one site.  We then show that it is possible to extend the computations to the case of the quantum field theory formulations \cite{GustafsonIsing}  that require more qubits but are guaranteed to scale efficiently for larger volume \cite{Lloyd1073}.

{\it 2. The model.}
We consider the transverse-field Ising model in one spatial dimension,
\begin{equation}
\label{eq:HamiltonianIsingFree}
\hat{H}_0 = - J \sum_{i=1}^{N - 1} \hat{\sigma}^x_i \hat{\sigma}^x_{i+1}  - h_T \sum_{i=1}^N \sigma^z_i.
\end{equation}
This model is very well understood \cite{RevModPhys.55.775} and discussed for NISQ devices  \cite{CerveraLierta2018exactisingmodel,Lamm:2018siq,GustafsonIsing,gustafson2019real}. It is equivalent to a theory of free fermions with subtle effects from the boundary.  Non-trivial interactions can be introduced with an extra term 
\begin{equation}
\label{eq:Hamiltonianint}
\hat{H}_{int} = u \sum_{i=1}^{N - 1} \hat{\sigma}^z_i \hat{\sigma}^z_{i+1}. 
\end{equation}

In order to perform Fourier transforms with a reduced number of qubits we first consider the quantum mechanics limit
$J \ll h_T$, where the model consists of energy bands that can be assigned a particle number. This amounts to neglecting terms of the form $\hat{\sigma}^+_i \hat{\sigma}^+_{i + 1}$ and their conjugates. For the reduced one-particle problem, the interaction introduces an effective potential at the end of the chain. This reduces the size of the Hilbert space from $2^N$ to $N$ and also allows approximate analytic calculations \cite{GustafsonIsing}. 

\begin{figure}[t]
\centering
    \includegraphics[width=0.35\textwidth]{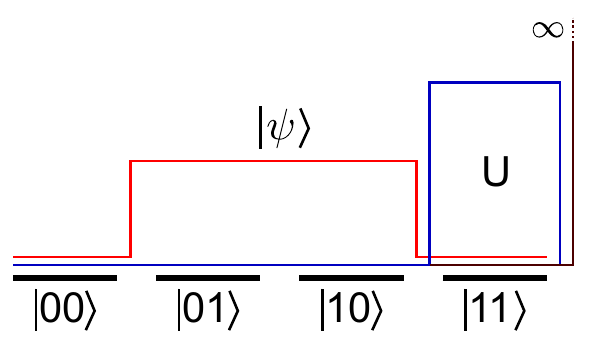}
    \caption{Visual depiction of the qubit states (black dashes), potential for the interacting (blue) and non-interacting case (dark red), and initial wave packet (light red). }
    \label{fig:setup}
\end{figure}

Specializing to the case  $N = 4$, we have, up to an unimportant additive constant, the following effective Hamiltonian matrix:
\begin{equation}
\label{eq:TensorDecomp}
\hat{H}_{eff} = 
\begin{pmatrix}
0 & -J & 0 & 0\\
-J & 0 & -J & 0 \\
0& -J & 0 & -J \\
0 & 0 & -J & U \\
\end{pmatrix}.
\end{equation}
This allows us to reduce the Hilbert space from four qubits needed for the $N=4$ field theory problem to two qubits for the one particle limit. The re-mapping is shown in Eq.(~\ref{2-to-4mapping}) with the correspondence illustrated in Fig. \ref{fig:setup}:
\begin{equation}
\label{2-to-4mapping}
\begin{split}
\dket{1000} &\rightarrow \dket{00}, ~\dket{0100} \rightarrow \dket{01}\\
\dket{0010} &\rightarrow \dket{10}, ~\dket{0001} \rightarrow \dket{11}.\\
\end{split}
\end{equation}
The Hamiltonian in Eq. (\ref{eq:TensorDecomp}) can now be written as 
\begin{equation}
    \label{eq:paulidecompham}
    \begin{split}
    \hat{H}_{eff} = & -J \sigma^x_{II} - \frac{J}{2}\Big(\sigma^x_I\sigma^x_{II} + \sigma^y_I\sigma^y_{II}\Big)\\
    &+ \frac{U}{4}(1 - \hat{\sigma}^z_I)(1 - \hat{\sigma}^z_{II}).
    \end{split}
\end{equation}
The subscripted roman numerals are used to indicate the use of our two-qubit decomposition.

{\it 3. Real-time scattering.}
The time delay $\Delta t_W$ of a wave packet with a sharply defined momentum $k$ is related to the derivative of the phase shift by the Wigner formula \cite{PhysRev.98.145}: 
\begin{equation}
\label{eq:we}
\Delta t_W=2\delta'(k)/(\partial E/\partial k),
\end{equation}
where $\partial E/\partial k$ is the group velocity, which in our case is $2J\sin(k)$.

We will explain that $\Delta t_W$ can be estimated from the first half of the real-time evolution of the scattering process and that it is actually twice the time delay illustrated in Fig. \ref{fig:diagram}.

We now report quantum computations for $J = 0.02$ and $U = 0.03$. In the 2-qubit Hilbert space, the momentum states for $k=\pm \pi/2$ read
\begin{equation}
    |k=\pm \pi/2\rangle=\frac{1}{2}(|00\rangle \pm i |01\rangle - |10\rangle \mp i |11\rangle).
\end{equation}
It is necessary for the initial wave packet to have some localization in space so that a distinct scattering event is visible, i.e., there is a point in time when the particle reaches the interaction region. As a side effect, the wave packet will have some momentum distribution because it no longer is a plane wave. 
As depicted in Fig. \ref{fig:setup}, we chose our initial wave packet to be $\dket{\pi/2}$ but restricted it to be non-zero only in the middle:
\begin{equation}
\label{eq:wp}
\dket{\psi} = \frac{1}{\sqrt{2}}\Big(\dket{01} + i\dket{10}\Big).
\end{equation}

We construct this wave packet on two sites with the following quantum circuit with all qubits initialized in the $\dket{0}$ state:
\begin{equation}
U_{\text{state prep}} = 
\begin{gathered}
\Qcircuit @C=-1.5em @R=-4em @! {
&\gate{X} & \multigate{1}{XX\Big(-\frac{3\pi}{2}\Big)} & \qw\\
&\qw      & \ghost{XX\Big(-\frac{3\pi}{2}\Big)} & \qw
}
\end{gathered}.
\end{equation}
The time evolution operator can be written as a combination of XX, YY, X, rotations and a controlled phase operation ($R_\phi$):
\begin{equation}
\begin{split}
&U_{\text{Tr}}(\rho, \theta) = \\
&\begin{gathered}
\Qcircuit @C=0em @R=-2em @!{
& \gate{R_x(2\rho)} & \multigate{1}{XX(\rho)} & \multigate{1}{YY(\rho)} & \ctrl{1} \\
& \qw             & \ghost{XX(\rho)} & \ghost{YY(\rho)} & \gate{R_{\phi}(\theta)} \\
}
\end{gathered},
\end{split}
\end{equation}
where $\rho = J \delta t$, $\theta = U \delta t$, and $\delta t = 12.5$. We use standard notations \cite{nielsen2000quantum} for the gates.
The very slow growth of the one-step error for large $\delta t$ \cite{ybook} allows us to reach $t=75$ with only six Trotter steps \cite{gustafson2019real}.

We then perform a quantum Fourier transform (QFTr) on these two qubits to take this state into momentum space:
\begin{equation}
U_{QFTr} =
\begin{gathered}
\Qcircuit @C=-1em @R=-2em @!{
& \qw & \gate{R_{\phi}(\pi/2)} & \gate{H} \\
& \gate{H} & \ctrl{-1} & \qw
}
\end{gathered}.
\end{equation}
After applying the QFTr, the qubit states $\dket{10}$ and $\dket{11}$ correspond to the momentum states $\dket{k}$ and \mbox{$\dket{-k}$} respectively, with $k=\pi/2$.

We define the probabilities to be in the $\dket{\pm k}$ state
\begin{equation}
P_{\pm}(t)\equiv|\dbra{\pm k} \psi(t)\rangle |^2,
\end{equation}
and their renormalized versions
\begin{equation}
\label{eq:rminus}
R_{\pm}(t)\equiv \frac{P_{\pm}(t)}{P_ + (t)+P_-(t)}
\end{equation}
which by design satisfy 
\begin{equation}
\label{eq:sumrule}
 R_+ + R_ -=1.   
\end{equation}
The real-time evolution provides the time $t^\star$ necessary to reach the symmetric situation where $P_+(t^\star)=P_- (t^\star)$. This corresponds to the time where a classical particle would hit the barrier. We can then compare $t^\star$ in the case where $U = 0$ and $U = 0.03$ and we call these times $t^\star_{free}$ 
and $t^\star_{int.}$ respectively. In order to determine the difference
\begin{equation}
\label{eq:dstar}
\Delta t^\star\equiv t^\star_{int.}-t^\star_{free} =\frac{\Delta t _W}{2}
\end{equation}
The relation to the time delay $\Delta t _W$ used in Eq. (\ref{eq:we}) is supported by numerical calculations provided later.  This relation can also be justified from the time-reversal argument that after $ t^\star_{int.}$ only {\it half} of the phase shift, $\delta(k)$, has built up while the other half builds after $t^\star$. This is why historically, the total phase shift is denoted $2\delta(k)$.

The time $t^\star$ is determined by the symmetric condition $R_-(t^\star)=R_+(t^\star)=0.5$.  This requires some continuous interpolation among the discrete values at the Trotter steps. We find that the sigmoid parametrization
\begin{equation}
R_{\pm}(t)\simeq 1/(1+\exp(\mp(\frac{ t-t^\star}{w}))),
\end{equation}
provides very good fits of the numerical data and satisfies the sum rule of Eq. (\ref{eq:sumrule}) because of its reflection property. This also implies that the time reversal with respect to $t^\star$ amounts to the momentum reversal. The parameter $w$ describes the width of the sigmoid.

\begin{figure}[ht!]
\vskip-3em
\includegraphics[width=0.5\textwidth]{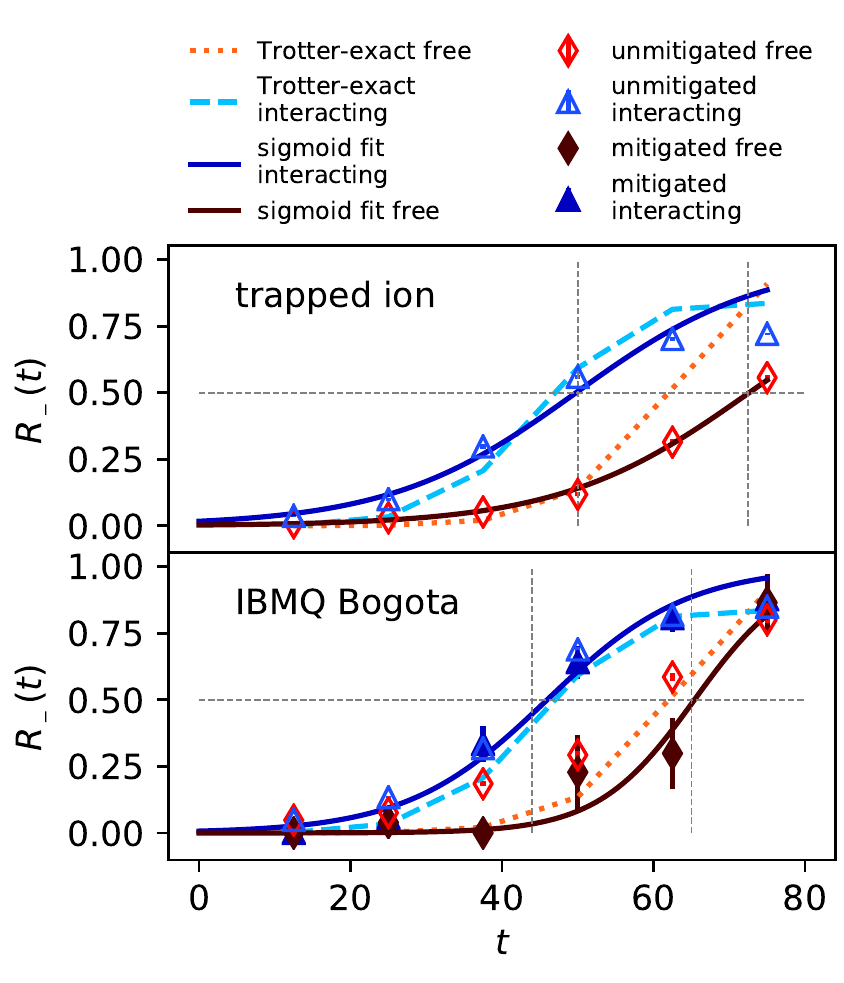}
\vskip-2em
\caption{Experimental results for $R_-(t)$ in the quantum mechanics limit versus time for a 4-site lattice using $J=0.02$, $h_T = 1.0$, and $U=0.03$ on two qubits for trapped ion (top), IBM (bottom) quantum computer.}
\label{fig:wallscatter}
\includegraphics[width=0.5\textwidth]{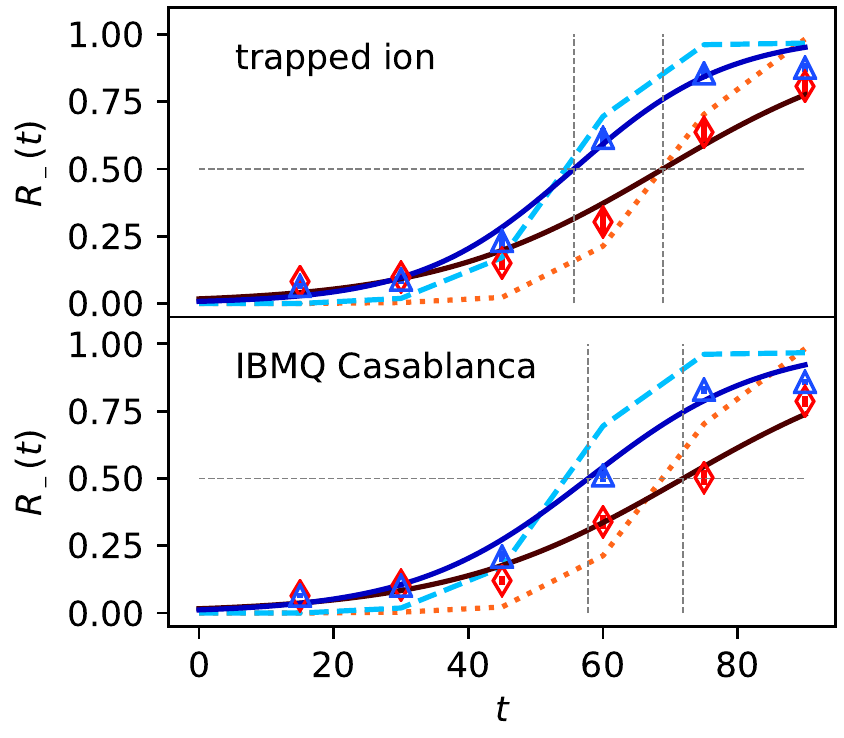}
\vskip-0em
\caption{Experimental results for $R_-(t)$ using the full Hamiltonian of Eq. (\ref{eq:HamiltonianIsingFree}) with and without the interaction term on four qubits for trapped ion (top), IBM (bottom) quantum computers. Only statistical errors are shown. }
\label{fig:qftevolution}
\end{figure}

The data for $R_-$ obtained from both quantum computers are shown in Fig. \ref{fig:wallscatter} together with the sigmoid fits where the values of $t^\star_{free}$ 
and $t^\star_{int.}$ are indicated by vertical lines crossing the fits at the 0.5 horizontal line. This provides the numerical values of $\Delta t^\star$ given in Table \ref{tab:tstar}. For comparison, we give the values obtained by doing sigmoid fits of the  continuous-time evolution (first column) and the Trotter steps (second column) calculated numerically at the same discrete times as the experimental data.  
The systematic errors are expected to be larger than the statistical errors and very difficult to estimate. Details of the fits are given in the Supplementary Material (SM). 

We see that both the IBM and trapped ion estimates provide larger absolute values of $\Delta t^*$ than the target values. This can be in part explained by the fact that the fits for the free process tend to lag below the Trotter steps for $t>50$ indicating a loss of coherence. 

Measurements from the IBMQ Bogota machine contain both the noisy data with just readout corrections and a mitigated version obtained using methods discussed in Ref. \cite{gustafson2019real} and which account for some slightly negative occupations at low $t$.
The trapped ion simulations include only readout corrections without noise mitigation (see SM).

We see that the quantum mechanics approximation allows us to perform the QFTr and get reasonable estimates of $\Delta t^*$, (Table \ref{tab:tstar}). We expect to improve the accuracy of these estimates in the near future.  The extension of this procedure for more than four sites requires an all-to-all connectivity and a CNOT depth increasing with the number of sites.  In contrast, the field theory calculation discussed in the next section, and which is our ultimate goal, requires more qubits but remains local \cite{Lloyd1073} with a constant CNOT depth.

\begin{table}[t!]
    \centering
    \begin{tabular}{|c|c|c|c|c|}
    \hline
         Type & Continuous & Trotter-exact & Trapped ions & IBM  \\ \hline 
         Q.M. & -17.5(1) & -13.7(9) & -23(3) & -20(3) \\ \hline
         Q.F.T. & -17(1) & -14.3(9) & -12(2) & -13(2) \\\hline
    \end{tabular}
    \caption{\label{tab:tstar}Results for $\Delta t^*$ in the quantum mechanics limit (Q.M) and full field theory (Q.F.T.) from sigmoid fits of the simulated continuous and  Trotter-exact  evolutions as well as the experimental data from the trapped ions and IBM quantum computers.}
    \vskip-2em
    \label{tab:phaseshifts}
\end{table}

{\it 4. Towards scalable field theory calculations.}
We have performed field theory computations with the same devices using four sites and four qubits.
This allows for shallower local circuits but requires a more expensive Fourier transform \cite{PhysRevLett.113.010401} as discussed in the SM. In Fig. \ref{fig:qftevolution} we show the results of these calculations on IBM's Casablanca and the Maryland ion trap machines using the full Hamiltonian with a Trotter step of $\delta t = 15$ with six Trotter steps. In both cases, unmitigated data is used for the analysis. Fig. \ref{fig:qftevolution} indicates that the discrepancies between the individual fits and the Trotter exact results are more pronounced than in the previous calculation. However the effects somehow compensate when we calculate the differences and we obtain time delays closer to the Trotter exact ones as shown in Table \ref{tab:tstar}.  
As discussed in the SM, the fit methods are similar to the previous ones. 
As improved quantum computing hardware platforms become available, we plan to use these upgraded facilities to get more accurate values of $\Delta t^\star$ and extend these calculations to 8 and 16 sites with one qubit per site.

\begin{figure}[t!]
\vskip-3em
\includegraphics[width=0.49\textwidth]{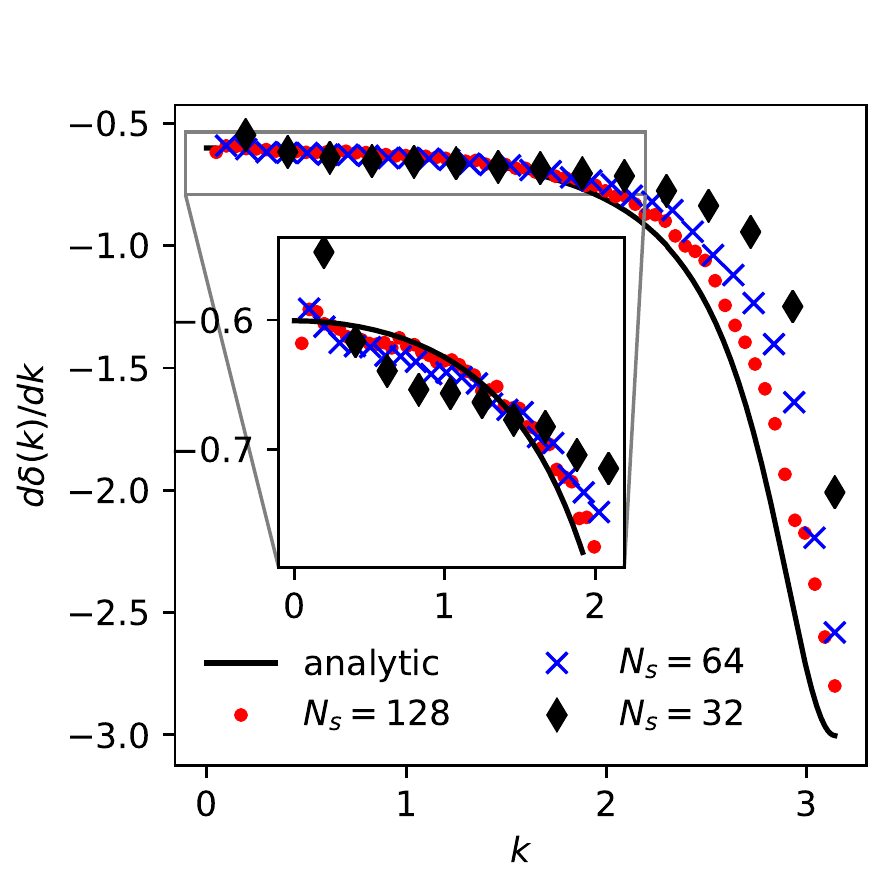}
\caption{Derivative of the phase shift with respect to momentum calculated numerically using the time delays and Eq. (\ref{eq:we}) for $N$= 32, 64 and 128 compared to the infinite volume results from Eq. (\ref{eq:dw}).}
\label{fig:wigner}
\end{figure}

{\it 5. Extracting the phase shifts from time delays.} 
We are now in position to discuss how to use Eq. (\ref{eq:we}) to estimate the phase shifts.  
First, we need to say that the wave packet in Eq. (\ref{eq:wp}) is not narrow and volume effects should be significant for $N=4$.
In order to estimate these effects we consider the one particle quantum mechanical problem at larger $N$ with narrow Gaussian wave packets. The details are given in the SM, where we also show that in the  infinite volume limit
\begin{equation}
\delta'(k)=-1+J\frac{U\cos(k)+J}{U^2+J^2+2JU\cos(k)}.
\label{eq:dw}
\end{equation}
The results at finite and infinite volume are 
compared in Fig. \ref{fig:wigner} illustrate the magnitude of the volume effects for $N$= 32, 64 and 128. 

Some integration procedure is needed in order to extract $\delta(k)$. In the quantum mechanics limit, we noticed that for small enough $k$, $\delta(k)$ could be approximated very well by $2E(k) \Delta t^\star$. It is shown in the SM that this feature can be explained from the Taylor expansion of the exact formula. This provides initial data that is reliable for $k\leq 0.5$. For larger $k$ an integration procedure based on Eq. (\ref{eq:we}) needs to be used.

{\it 6. Conclusions.}
We have proposed a method to estimate the time delay using the early steps of the real-time evolution.  We have given a proof of principle that actual computations of the time delays can be implemented on both the IBM superconducting transmon and trapped ion hardware platforms for a quantum mechanics limit with two qubits and the field theory formulation with four qubits.  
There is plenty of room for optimization for both devices and we do not claim that our results allow a systematic comparison between the devices.  We expect that the field theory calculations should feasible for a larger number of qubits in the near future. A detailed comparison with existing real-time methods in one spatial dimension \cite{vidal2004,white2004,verstraete2004,daley2004} would be of great interest. Quantum computations for quantum Ising models in two spatial dimensions 
could offer the possibility to reach quantum advantage. 

\vskip20pt
\begin{acknowledgments}
{\it Acknowledgments}. E. Gustafson and Y. Meurice were supported in part by the U.S. Department of Energy (DoE) under Award Number DE-SC0019139. They thank the members of QuLAT for suggestions and comments. We also thank Wayne Polyzou and the math physics seminar at the University of Iowa for fruitful discussions. P. Dreher was supported in part by the U.S. Department of Energy (DoE) under award DE-AC05-00OR22725. We thank North Carolina State University (NCSU) for access to the IBM Q Network quantum computing hardware platforms through the NCSU IBM Q Hub. N. M. Linke acknowledges support by the DoE NP Quantum Horizons program, award number DESC0021143, and the NSF grant no. PHY-1430094 to the PFC@JQI. We are extremely grateful to Alaina M. Green, Cinthia Huerta Alderete and Nhung H. Nguyen for help with trapped-ion measurements. 
\end{acknowledgments}



\pagebreak
\newpage
\begin{center}
\textbf{\large Supplemental Materials}
\end{center}
\setcounter{equation}{0}
\setcounter{figure}{0}
\setcounter{table}{0}
\setcounter{page}{1}
\renewcommand{\theequation}{S\arabic{equation}}
\renewcommand{\thefigure}{S\arabic{figure}}
\subsection{Quantum Computing Hardware Platforms}
This project used both the IBM Quantum Network superconducting transmon hardware platforms and a trapped ion machine.  For the IBM quantum computers, the project used IBM$_Q$ Bogota (5 qubit) and IBM$_Q$ Casablanca (7 qubit) machines. Both IBM platforms are rated with a quantum volume of $32$.  Access to these machines was through the IBM cloud via the NC State IBM Quantum Hub.

The ion trap quantum computer used in this study consists of a chain of $^{171}$Yb$^+$ ions confined in a linear Paul trap and laser-cooled to their motional ground states \cite{Debnath2016,landsman_verified_2019}. The qubits are defined in the hyperfine-split $^2S_{1/2}$ manifold as $|0\rangle=|F=0, m_F=0\rangle,|1\rangle=|F=1,m_F=0\rangle$, with a splitting of $12.642821\:$GHz and are insensitive to magnetic field fluctuations to first order. Qubits are initialized to the $|0\rangle$ state by optical pumping \cite{olmschenk_manipulation_2007}. Coherent operations are achieved by illuminating the ions with a pair of counter-propagating Raman beams at 355-nm which have a beat-note that is set resonant with the qubits or near the motional sidebands of the chain \cite{islam_beat_2014}. One of the Raman beams illuminates the entire chain. The other one is split into individual beams that are each matched onto one channel on a multi-channel acoustic-optic modulator (AOM). The amplitude, frequency, and phase of each beam can be modulated independently by the corresponding RF signal. Individual addressing is achieved by focusing each of these beams onto one single ion. Each ion is also matched onto a distinct channel of a photo-multiplier tube (PMT) that collects state-dependent fluorescence for individual state readout \cite{olmschenk_manipulation_2007}. There are two mechanisms of quantum control. Single qubit rotations are performed by driving Rabi rotations of the target ion on resonance.  Two-qubit entanglement is implemented by creating an effective Ising spin-to-spin interaction via transient entanglement between the qubits and the motional modes with the Raman beat-note tuned near the motional sidebands \cite{milburn_ion_2000,molmer_multiparticle_1999,solano_deterministic_1999}. Only qubit-to-qubit entanglement remains after the motional modes are dis-entangled at the end of the scheme \cite{choi_optimal_2014}. Entanglement can be created between any pair of ions owing to the long-range Coulomb interaction \cite{linke_experimental_2017}. Typical gate fidelities are around $99.5(2)\%$ for the single-qubit gates and 98(1)\% for the two-qubit entangling gates. The main sources of two-qubit gate errors are residual entanglement between the motional modes and the qubits due to laser intensity fluctuations and motional heating. State preparation and measurement (SPAM) errors are accounted for by applying the inverse of an independently measured state-to-state error matrix when analyzing the data.

\subsection{Determining time delay with sigmoids}
\begin{widetext}
\begin{table}[b]
\centering
\begin{tabular}{|c|c|c|c|c|c|}
\hline
               Type &         $t*_{free}$ &          $t*_{int}$ &            $\Delta t*$ &        $w_{free}$ &         $w_{int}$\\\hline\hline
  Continuous Sampled &    68.1(1) &   50.6(2) &    -17.5(1) &   6.47(8) &   7.12(2)\\
           Trotter &    61.8(6) &    48.2(7) &    -13.7(9) &    5.8(4) &    8.8(6)\\
            IBM un-mitigated & 57.9(8) & 46(1) & -12(1) & 13.6(5) & 12.0(7)\\
         IBM mitigated & 65(3) & 47(1) & -19(3) & 6(2) & 10.4(7) \\
            Trapped Ion & 72.6(6) & 49(2) & -22(2) & 12.5(4) & 12(1) \\\hline
\end{tabular}
\caption{Result of sigmoid fits to a sampling of the exact evolution operator (continuous), Suzuki-Trotterization, the un-mitigated and mitigated data from IBM hardware, and the University of Maryland Trapped Ion quantum computer (Quantum Mechanics limit).}
\label{tab:sigmoidfits}
\centering
\begin{tabular}{|c|c|c|c|c|c|}
\hline
               Type &         $t*_{free}$ &          $t*_{int}$ &            $\Delta t*$ &        $w_{free}$ &         $w_{int}$ \\\hline\hline
  Continuous Sampled &    68.0(7) &   50.9(8) &    -17(1) &   5.25(3) &   8.2(2)\\
           Trotter &    69.1(6) &    54.8(7) &    -14.3(9) &    6.8(6) &    6.5(4)\\
          IBM un-mitigated & 72(2) & 59(1) & -13(2) & 18(1) & 14.2(9)\\
          Trapped Ions & 69(1) & 57(1) & -12(2) & 17(1) & 13(1) \\\hline
\end{tabular}
\caption{Result of sigmoid fits to a sampling of the exact evolution operator (continuous), Suzuki-Trotterization, the un-mitigated from IBM hardware, and the University of Maryland Trapped Ion quantum computer  (QFT).}
\label{tab:sigmoidfitsqft}
\end{table}
\end{widetext}
Using the sigmoid for the fit of the quantities $R_-(t)$ we find the time delay.
We have separated out each of the various data sets into lines. The continuous sampling uses 100 equally spaced time steps between $t = 0$ and $t = 75$ with a corresponding uncertainty of $0.001$. Similarly we also used 6 equally spaced data points sampled from the continuous distribution at time which are multiples of $t = 12.5$ with the same flat error rate of $0.001$. The Trotter-exact was sampled at times steps of $\delta t = 12.5$ and had an uncertainty of $\delta P_{\pm}(t) = \sqrt{P_{\pm} - P_{\pm}^2(t)} / \sqrt{1000}$. The remaining three lines were sampled by fitting to the first five data points for the interacting case ($int$) and six data points for the free case. This was done to make sure that the value of $R_{-}(t)$ is at least $0.5$ and that the system has not reached an asymptotic state. Stated in a different way, 
in the interacting case, the signal flattens below 1 after 5 Trotter steps due to the reflection on the left wall. The last data point was not used for the fit and is always below the fit. 

The systematic errors are significant and very difficult to estimate. It is possible to get an idea of their magnitude by looking at the first time step where 
a detailed examination  shows that the measured values of $R_-$ are significantly larger than the Trotter exact values and the estimated statistical errors. In absence of a reliable model for the 
systematic errors, we estimated the error on the parameters using the jackknife method. 

The fitting methods for the field theory results are similar, the only difference being a slightly larger Trotter step.

\subsection{QFT circuit}
Initially we want to scale up from the quantum mechanics picture to the field theory perspective, in other words the full Ising model. As in the quantum mechanical case, we will need to break up this circuit into three parts: 1)
 state Preparation, 2)
    Trotter Evolution, 3)
    momentum Projection.
The various circuits and the relative costs are discussed below. 

We want to prepare the same initial state that we proposed in the main text mapped on to the Ising Model.
This initial state that we prepared:
\begin{equation}
    \label{eq:quantummechanicspacket}
    \dket{\psi_i} = \frac{1}{\sqrt{2}}\Big(\dket{01} + i \dket{10}\Big)
\end{equation}
can be remapped to the Ising model by using the inverse mapping originally described in the text:
\begin{equation}
\begin{split}
\dket{00} &\rightarrow \dket{1000}, ~\dket{01} \rightarrow \dket{0100}\\
\dket{10} &\rightarrow \dket{0010}, ~\dket{11} \rightarrow \dket{0001}.\\
\end{split}
\end{equation}

This remapping will give us the initial state:
\begin{equation}
    \label{eq:initialstatefieldtheory}
    \dket{\psi_i} = \frac{1}{\sqrt{2}}\Big(\dket{0100} + i \dket{0010}\Big).
\end{equation}
This state can easily be prepared with 1 $XX$ gate and one single qubit rotation.
\begin{equation}
\begin{gathered}
    \Qcircuit{
    \lstick{\dket{0}} & \qw      & \qw                      & \qw \\
    \lstick{\dket{0}} & \gate{X} & \multigate{1}{XX(-\pi/2)} & \qw \\
    \lstick{\dket{0}} & \qw      & \ghost{XX(-\pi/2)}        & \qw \\
    \lstick{\dket{0}} & \qw      & \qw                      & \qw \\
    }
    \end{gathered} = \hat{U}_{prep}
\end{equation}

The time evolution of the system can be easily written in terms of a circuit requiring at most 3 $XX$ gates per Trotter step when using a four site system and this will be 2 $XX$ gates deep. This circuit will end up being an application of entangling rotations followed by a set of single qubit phasing operations. 
\begin{widetext}
\begin{equation}
    \hat{U}_{Trot}(\delta t; J, U, h_T) = 
    \begin{gathered}
    \Qcircuit{
    &\qw & \gate{R_z(2h_T\delta t )}        & \multigate{1}{XX(2 J \delta t)} & \qw & \qw \\
    &\qw & \gate{R_z(2h_T\delta t )}        & \ghost{XX(2 J \delta t)} & \multigate{1}{XX(2 J \delta t)} & \qw \\
    &\qw & \gate{R_z(2h_T\delta t )}        & \multigate{1}{XX(2 J \delta t)} & \ghost{XX(2 J \delta t)} & \qw \\
    &\qw & \gate{R_z(2(h_T + U / 2)\delta t)} & \ghost{XX(2 J \delta t)} & \qw & \qw \\
    }
    \end{gathered}
\end{equation}
\end{widetext}

The final portion of the quantum simulation is a Fourier transformation to take the circuit into momentum space \cite{Ferris_2014, Kivlichan_2020}. This transformation can easily be done using four two-qubit operations. 
\begin{widetext}
\begin{equation}
    \label{eq:fastfourier}
    \hat{U}_{fft} = 
    \begin{gathered}
    \Qcircuit{
    &\qw & \qw & \sgate{F}{1} & \gate{R_{\phi}(\pi/2)} & \qw & \sgate{F}{2} & \qw \\
    &\qw & \ctrl{1} & \gate{F}  & \ctrl{1} & \sgate{F}{2} &\qw & \qw \\
    &\qw & \gate{Z} & \sgate{F}{1} &  \gate{Z} & \qw & \gate{F} &\qw \\
    &\qw & \qw & \gate{F} & 
    \qw & \gate{F} & \qw & \qw
    }
    \end{gathered}
\end{equation}
\end{widetext}
Where $F$ is given by the matrix:
\begin{equation}
    F = 
    \begin{pmatrix}
    1 & 0 & 0 & 0 \\
    0 & \frac{1}{\sqrt{2}} & \frac{1}{\sqrt{2}} & 0 \\
    0 & \frac{1}{\sqrt{2}} & -\frac{1}{\sqrt{2}} & 0 \\
    0 & 0 & 0 & -1 \\
    \end{pmatrix}.
\end{equation}
This can be decomposed into at worst 4 CNOT gates. It should be noted this is different from the traditional quantum Fourier transform which uses binary encoding for the qubits.

The measurement is done then in the z-basis and we select out the states $|0010\rangle$ and $|0001\rangle$ for the $|k\rangle$ and $|-k\rangle$ state respectively.
\subsection{Phase Shift Derivation}

The discrete Schroedinger equation obtained in the limit of small $J$, with $U=0$, and no boundaries admits plane wave solutions $e^{\pm i kx}$ with energy $2J(1-\cos(k))$. 
The additive constant has been adjusted in order to have zero energy at zero momentum.  
If we introduce a right wall and impose $\psi(N+1)=0$, we need a relative phase $-e^{ i 2k(N+1)}$ for the reflected wave $e^{- i kx}$. 
If, in addition, we also introduce a repulsive potential $U>0$ at the rightmost site $N$, the problem with a right wall can be solved by introducing the phase shift. A short calculation shows that: 
\begin{equation}
e^{i2\delta(k)}=e^{-i2k}\frac{U+Je^{ik}}{U+Je^{-ik}},
\label{eq:phaseshift}
\end{equation}

We are now in position to make two independent calculations of $\Delta t^\star$. In the first one, we use the general Eqs. (\ref{eq:we}) and (\ref{eq:dstar}) together with the exact expression for $\delta'(k)$ from Eq. (\ref{eq:phaseshift}) to get:
\begin{equation}
\Delta t^*=(-1+J\frac{U\cos(k)+J}{U^2+J^2+2JU\cos(k)})/2J\sin(k).
\label{eq:w2}
\end{equation}
The other method consists in using the real-time evolution to calculate $R_-(t)$ and determine the time $t^\star$ where $R_-(t^\star)=0.5$. This second method depends on the volume and the details of the initial wavepacket.
It is important to prepare a wave packet that is broad enough in space to have sharply defined momentum but not broad enough to ``bounce back on the left wall" too early. In the numerical calculations, we used an initial wavepacket
\begin{equation}
\psi(x)\propto e^{ikx-(x-N/2)^2/(N/4)^2}.
\label{eq:wavepacket}
\end{equation}
 The comparison between these results and Wigner formula are shown in the text.

A few words of explanation regarding the low $k$ approximate formula for $\delta(k)$. 
From Eq. (\ref{eq:phaseshift}),
we can expand $\delta(k)$ in power of $k$. 
For our numerical values $J=0.02$ and $U=0.03$, we obtain
\begin{equation}
\delta(k)\simeq -0.6k+0.008k^3+\dots
\end{equation}
As we see,  for $0\leq k\leq 0.5$ the correction to the linear approximation are less then one percent and we can in good approximation use 
\begin{equation}
\delta(k)\simeq \delta '(k) k.
\end{equation}


\bibliography{bibliographyscattering.bbl}

\end{document}